# Title page:

## PINN-based short-term forecasting of fault slip evolution during the 2010 slow slip event in the Bungo Channel, Japan

**Masayuki Kano**[1,2] (ORCID: 0000-0002-7288-4760), **and Rikuto Fukushima**[3] (ORCID: 0009-0007-0463-1458)

- Masayuki Kano, Disaster Prevention Research Institute, Kyoto University, Japan, kano.masayuki.4m@kyoto-u.ac.jp
- Rikuto Fukushima, Department of Geophysics, Stanford University, U.S., rfukushima@stanford.edu

**Corresponding author: Masayuki Kano, kano.masayuki.4m@kyoto-u.ac.jp**

# Abstract

Monitoring and forecasting fault slip evolution are fundamental for understanding earthquake cycles and assessing future seismic hazards. This study proposes a physics-based data assimilation framework that integrates geodetic observations with fault mechanics introducing spatial heterogeneity in frictional properties, with a particular focus on short-term fault slip forecasting. The proposed method employs physics-informed neural networks (PINNs) to calculate fault slip evolutions and to optimize the spatial distribution of frictional properties and is applied to the 2010 slow slip event beneath the Bungo Channel, southwest Japan, by changing the data period to be assimilated. When observation data reflecting the initial phase of slip acceleration is assimilated, a velocity-weakening frictional region is inferred beneath southwest Shikoku, corresponding to the initial nucleation area of the slow slip event. Using this frictionally heterogeneous model, we demonstrate that the PINN-based data assimilation framework successfully forecasts slow transient slip even when observation data corresponding only to slip acceleration are assimilated, whereas forecasts based on frictionally homogeneous models result in unstable fast slip. This difference can be interpreted as a consequence of introducing frictional heterogeneity, which allows both the characteristic size of the slipping region and the critical nucleation size to be variable, leading to stable slip evolution consistent with observations. When longer observation periods are assimilated, a velocity-strengthening region emerges around the velocity-weakening patch, progressively restricting the direction of slip propagation. This velocity-strengthening region can be interpreted as a mechanical constraint within the adopted model framework, linking the slip regions required to reproduce the observed geodetic time series. These results highlight the capability of PINN-based data assimilation incorporating geodetic observations and fault mechanics. The proposed framework provides a promising approach toward more realistic fault slip monitoring and short-term forecasting, and offers a foundation for future extensions incorporating uncertainty quantification and probabilistic forecasting.

# Main Text

## Introduction

A wide spectrum of fault slip phenomena, including coseismic fast slip, afterslip, slow slip events (SSEs), and steady plate motion, has been observed in plate subduction zones. Monitoring these diverse fault slip processes provides a comprehensive understanding of the entire earthquake cycle, from interseismic strain accumulation to coseismic rupture and postseismic relaxation. Among these phenomena, SSEs, which are the primary focus of this study, are characterized by slow, transient fault motions with durations ranging from days to years, in contrast to rapid coseismic slip (e.g., Obara and Kato 2016). Although SSEs do not generate strong ground motions, seismic and geodetic evidence suggests that SSEs can play an important role in earthquake preparation processes: SSEs have been inferred to precede megathrust earthquakes (e.g., Kato et al. 2012; Ruiz et al. 2014), implying that SSEs may modulate stress conditions on adjacent locked fault segments. Therefore, continuous monitoring and quantitative characterization of SSEs are expected to provide valuable information for evaluating future slip potential. However, observational evidence alone is generally insufficient to forecast the future evolution of ongoing SSEs. This limitation motivates the development of physics-based forecasting frameworks that can integrate observations with fault mechanics to infer and forecast SSE evolution.

This study particularly focuses on short-term forecasting of SSEs using a physics-based approach. Data assimilation (DA), which dynamically integrates observation data with numerical modeling governed by physical laws, provides a promising framework for such physics-based forecasting (e.g., Lewis et al. 2006; Fletcher 2022). DA techniques were originally developed in meteorology and have been widely implemented in operational weather and ocean forecasting systems. More recently, a wide variety of DA methods have been proposed for fault slip monitoring problems, including Markov chain Monte Carlo (MCMC) method (Fukuda et al. 2009; Kano et al. 2024), ensemble Kalman filter (Hirahara and Nishikiori 2019; van Dinther et al. 2019; Diab-Montero et al. 2023), particle filter (Mitsui et al. 2010; Hori et al. 2014), adaptive Gaussian mixture filter (Diab-Montero et al. 2025), particle flow filter (Diab-Montero et al. 2025), and adjoint method (Kano et al. 2013; 2015; Ohtani et al. 2025). Most of these studies have primarily focused on developing DA methodologies to optimize fault slip properties, i.e., frictional parameters that control fault slip behavior. As a result, these approaches have largely been validated through numerical experiments rather than real observation data.

Kano et al. (2024, hereafter K24) were the first to apply DA to real SSE observations recorded during the 2010 Bungo Channel SSE in southwest Japan (Figure 1a). In K24, the subducting plate interface was modeled as a planar rectangular fault consisting of a single circular SSE patch, originally adopted by Hirahara and Nishikiori (2019). K24 estimated spatially uniform frictional parameters within the SSE patch while fixing its geometry. Despite these strong constraints, they demonstrated that such a simplified fault model could reproduce the long-term trend of the observed SSE signal.

K24 further investigated the short-term forecasting capability of their simplified model during an ongoing SSE. They demonstrated that assimilating observations corresponding to the slip acceleration phase alone could lead to a forecast of fast slips without imposing any prior constraints on the frictional parameters to restrict the solution to SSEs. This behavior was interpreted as resulting from the critical nucleation size, $R_c$, calculated from the estimated frictional parameters being smaller than the fixed radius of the SSE patch, $R$, thereby leading to unstable slip (Chen and Lapusta 2009).

Subsequently, Fukushima et al. (2025) relaxed the strong geometric and frictional constraints imposed in K24 by allowing for spatially heterogeneous frictional properties on the fault. Although estimating spatially variable frictional parameters requires solving a high-dimensional optimization problem, they showed that an emerging machine-learning approach — physics-informed neural networks (PINN; Raissi et al. 2019) — can successfully address this challenge. Through numerical experiments, Fukushima et al. (2025) showed that PINN can reconstruct a spatially heterogeneous SSE fault patch.

Building on this work, Fukushima et al. (2026) applied the PINN-based DA framework to the real SSE observations used in K24 and revealed an ellipsoidal SSE patch beneath the Bungo Channel, achieving a quantitatively improved fit to geodetic observations. To our knowledge, their fault slip model represents the first quantitative result that simultaneously explains observed surface deformation and satisfies the fault mechanics in terms of fault slip monitoring. As a next step, the present study extends their PINN-based DA approach to short-term forecasting of fault slip, and investigates how the introduction of frictional heterogeneity influences slip forecast performance.

## PINN-based DA framework

PINNs incorporate observation data into physics-based models using neural networks (NNs). This section summarizes the PINN-based DA framework adopted in this study, including the Global Navigation Satellite System (GNSS) observations and the Bungo Channel SSEs, the governing equations and fault model, and the problem settings of the PINN. To enable a direct comparison with the DA results of K24, we use the same GNSS data and fault model as those adopted in K24.

### *GNSS observations and the Bungo Channel SSEs*

Geodetic observations have revealed that SSEs with durations of approximately 1.0 yr and moment magnitude of ~6.8–6.9 repeatedly occurred beneath the Bungo Channel in southwest Japan, where the Philippine Sea Plate subducts beneath the Amurian Plate (Figure 1a) (Hirose et al. 1999; Kobayashi and Yamamoto 2011; Ozawa et al. 2013; Yoshioka et al., 2015; Seshimo and Yoshioka 2022). Crustal deformation associated with these SSEs has been continuously observed by the GNSS Earth Observation Network System (GEONET), operated by the Geospatial Information Authority of Japan. We used GNSS data from 86 GEONET stations (Figure 1a), consistent with K24. The data were preprocessed using the GipsyX-1.4 software (Bertiger et al. 2020) to obtain daily position time series. This dataset is largely identical to that used in K24 (see K24 for details of the preprocessing procedure), except that the data period is from 2006.5 to 2012.5, which is one year longer than in K24 and consistent with Fukushima et al. (2026). Linear trends during the inter-SSE period were estimated from the first two years (2006.5-2008.5) and removed from the entire time series. Figure 1b shows an example of the GNSS displacement time series at station 1059, which is located near the SSE source region. The SSE initiated around 2010 beneath southwest Shikoku and subsequently propagated westwards, lasting until approximately 2011 (Yoshioka et al. 2015; Fukushima et al. 2026). In the following analysis, we assimilate GNSS data using six different data periods starting from 2006.5 and ending at 2012.5, 2011.5, 2011.0, 2010.5, 2010.0, and 2009.5, respectively. The DA result using the full data from 2006.5 to 2012.5 corresponds to the case presented in Fukushima et al. (2026).

*Physics-based model of the Bungo Channel SSEs*

To calculate spatio-temporal evolution of SSEs, we adopt the fault model in the Bungo Channel originally introduced by Hirahara and Nishikiori (2019) and subsequently used in several studies (Kano et al. 2024; Ohtani et al. 2025; Fukushima et al. 2025; 2026). In this model, the subducting plate interface is represented as a single planar rectangle fault with a dip angle of 15° and dimensions of 120 km × 100 km in the strike and dip directions, respectively. The center of the fault plane is located at a depth of 25 km. The entire fault is then subdivided into 2 km × 2 km subfaults, resulting in 3,000 subfaults. In each subfault, fault slip evolution is assumed to obey the following governing equations:

$$\tau_i(t) = \sum_j k_{ij}(v_{\mathrm{pl}}t - s_j(t)) - \frac{G}{2\beta}v_i(t), \tag{1}$$

$$\tau_i(t) = \tau_{0i} + a_i\sigma \ln\left(\frac{v_i(t)}{v_0}\right) + b_i\sigma \ln\left(\frac{v_0\theta_i(t)}{L_i}\right), \tag{2}$$

$$\frac{d\theta_i(t)}{dt} = 1 - \frac{v_i(t)\theta_i(t)}{L_i}. \tag{3}$$

Equation (1) represents the quasi-dynamic equation of motion with radiation damping (Rice 1993), while Equation (2) describes the rate- and state- dependent friction law with the aging law for state evolution given by Equation (3) (Dieterich 1979; Ruina 1983). In these equations, $\tau_i(t)$, $s_i(t)$, $v_i(t)$ (= $\mathrm{d}s_i(t)/\mathrm{d}t$), and $\theta_i(t)$, denote the shear stress, slip, slip velocity, and state variables on subfault $i$ at time $t$, respectively. The slip response function $k_{ij}$ represents the shear stress change on subfault $i$ due to a unit slip on subfault $j$ and is calculated assuming an isotropic, homogeneous elastic half-space (Okada 1992). The slip direction is assumed to be parallel to the relative plate motion of the subducting Philippine Sea plate to the Amurian plate (Sella et al. 2002). The plate velocity $v_{\mathrm{pl}}$, shear modulus $G$, and shear wave velocity $\beta$ are set to be 6.5 cm/yr (Miyazaki and Heki 2001), 40 GPa, and 3.0 km/s, respectively. The frictional parameters $a_i$, $b_i$, and $L_i$ control the fault slip behavior on each subfault. Slip stability is governed by the characteristic fault patch size $R$ and the critical nucleation size $R_c$, defined in the following relationship (Chen and Lapusta 2009):

$$R_{\mathrm{c}} = \frac{\pi}{4}\frac{GbL}{\sigma(b-a)^2}. \tag{4}$$

When $R > Rc$, fault slip becomes unstable and can result in seismic slip, whereas $R < Rc$ leads to more stable slip. For example, Hirahara and Nishikiori (2019) successfully reproduced SSEs by prescribing a circular SSE patch with a radius of $R$ = 35 km and appropriate frictional parameters corresponding to $R/R\mathrm{c} \sim 0.47$. The reference shear stress $\tau_0$ in Equation (2) corresponds to the shear stress for steady-state sliding at a reference velocity $v_0$, and the effective normal stress $\sigma$ is set to be 100 MPa in this study. By combining Equations (1)–(3), we obtain an ordinary differential equation governing the temporal evolution of slip velocities, which can be written in the following form:

$$\frac{dv_i(t)}{dt} = \frac{\sum_j k_{ij}(v_{\mathrm{pl}} - v_j(t)) - \frac{b_i\sigma}{\theta_i(t)}\left(1 - \frac{v_i(t)\theta_i(t)}{L_i}\right)}{\frac{a_i\sigma}{v_i} + \frac{G}{2\beta}}. \quad (5)$$

*Problem setting of PINN-based DA*

PINNs provide a flexible framework for solving both forward and inverse problems involving differential equations with a set of model parameters (Raissi et al. 2019), and have more recently been applied to inverse problems in seismology by integrating physics-based equations with observation data (e.g., Waheed et al. 2021; Rasht-Behesht et al. 2022; Rucker and Erickson 2024). In the present study, we employ PINNs for fault slip forecasting and frictional parameter optimization, following the framework of Fukushima et al. (2026). Below, we briefly summarize the PINN-based DA framework (see Fukushima et al. 2026 for details).

First, we introduce two fully-connected NNs to represent simulation variables and the spatial distributions of frictional parameters. The first network, NN_A, takes time $t$ and the two-dimensional fault locations $(x, y)$ as inputs and outputs the slip velocity $v(t, x, y)$ and the state variable $\theta(t, x, y)$ at the corresponding time and location. The second network, NN_B, takes the spatial location $(x, y)$ as inputs and outputs the corresponding frictional parameters *a*- $b(x, y)$, $a(x, y)$ and $L(x, y)$. We adopt the same network architectures as those in Fukushima et al. (2026).

We then define the loss function $J$ as the sum of the model residuals $J_{\mathrm{ode}}$, the data misfit term $J_{\mathrm{data}}$, and loss for initial condition $J_{\mathrm{ini}}$:

$$J = J_{\mathrm{ode}} + w_{\mathrm{data}} J_{\mathrm{data}} + w_{\mathrm{ini}} J_{\mathrm{ini}}, \tag{6}$$

$$J_{\mathrm{ode}} = \frac{t^{*2}}{N_{ct} N_{cxy}} \sum_{i=1}^{Nct} \sum_{j=1}^{Ncxy} \left( \overline{r_v}(t_i, x_j, y_j)^2 + \overline{r_\theta}(t_i, x_j, y_j)^2 \right), \tag{7}$$

$$\overline{r_v}(t_i, x_j, y_j) = \frac{1}{v_{\mathrm{NN}}} \left( \frac{dv_{NN}}{dt} - f(v_{\mathrm{NN}}, \theta_{\mathrm{NN}}) \right), \tag{8}$$

$$\overline{r_\theta}(t_i, x_j, y_j) = \frac{1}{\theta_{\mathrm{NN}}} \left( \frac{d\theta_{NN}}{dt} - g(v_{\mathrm{NN}}, \theta_{\mathrm{NN}}) \right), \tag{9}$$

$$J_{\mathrm{data}} = \frac{1}{N_{\mathrm{data}} N_t} \sum_{i=1}^{Nt} \left( (\mathbf{H}\mathbf{s}_{\mathrm{NN}}(t_i) - \mathbf{d}_i)^{\mathrm{T}} \mathbf{R}^{-1} (\mathbf{H}\mathbf{s}_{\mathrm{NN}}(t_i) - \mathbf{d}_i) \right), \tag{10}$$

$$J_{\mathrm{ini}} = \frac{1}{N_{cxy}} \sum_{i=1}^{Ncxy} \left( \log \frac{v_{\mathrm{NN}}(0, x_i, y_i) \theta_{\mathrm{NN}}(0, x_i, y_i)}{L_{prior}} \right)^2 \tag{11}$$

The weighting parameters are set to be $w_{\mathrm{data}} = 10^{-3}$ and $w_{\mathrm{ini}} = 10^{-4}$ following Fukushima et al. (2026). Equations (8) and (9) represent the residuals of the ordinary differential equations corresponding to Equations (5) and (3), respectively. We denote the right-hand sides of Equations (5) and (3) by $f(v,\theta)$ and $g(v,\theta)$, respectively, and $v_{\mathrm{NN}}$ and $\theta_{\mathrm{NN}}$ represent the outputs of NN_A. $N_{ct}$ and $N_{cxy}$ denote the numbers of collocation points in time and space, respectively, at which the residuals are evaluated. The collocation points are uniformly distributed in both time and space, with $N_{ct} = 219$ and $N_{cxy} = 3{,}000$. These spatial and temporal distributions of collocation points are adequate to capture the essential features of SSE evolution. Specifically, the collocation points are distributed over the entire time domain (2006.5-2012.5), regardless of the length of the assimilation period. No observation data are assimilated in the forecast period; therefore, the temporal evolution of fault slip in this period is determined solely by the governing equations and the simultaneously optimized frictional parameters. In this sense, the forecast corresponds to a physics-based extrapolation beyond the assimilation period. The term $J_{\mathrm{ode}}$ is non-dimensionalized using the characteristic time $t^*$ defined as $L_{\mathrm{prior}} / v_{\mathrm{pl}}$ (Segall 2010), where $L_{\mathrm{prior}}$ is set to be 4.0 cm following Hirahara and Nishikiori (2019). Equation (10) represents the data misfit between the calculated and the observed displacements. Here, $\mathbf{d}_i$ consists of the observed displacements at the $i$-th epoch $t_i$ for all GNSS stations, with a total length of $N_{\mathrm{data}}$, and $N_{\mathrm{t}}$ is the total number of epochs. The variable $\mathbf{s}_{\mathrm{NN}}$ represents the slip at all subfaults, numerically integrated from $v_{\mathrm{NN}}$, and $\mathbf{H}$ is the observation matrix that projects fault slip onto surface displacements at each GNSS station, calculated assuming an isotropic, homogeneous elastic half-space (Okada 1992). The covariance matrix for observations $\mathbf{R}$ is set to be identical to that adopted in K24. Equation (11) imposes a constraint that the fault stress state at the initial epoch is close to steady-state conditions, which represents prior information for regularizing inversion.

The loss function (Equation (6)) is minimized using the L-BFGS method (Liu and Nocedal 1989) until convergence or until the maximum number of iterations (40,000) is reached. To facilitate NN parameter optimization, we initialize the NN parameters using a transfer learning approach following Fukushima et al. (2026), in which the networks are pretrained using synthetic data generated based on the K24 results. The optimized NN parameters obtained at the final iteration are then used to calculate the spatio-temporal evolution of fault slip velocities, the corresponding surface deformation, and the spatial distribution of frictional parameters. We confirm that, in the following results, the model residuals $J_{\mathrm{ode}}$ are well converged. This is demonstrated by calculating the displacement time series using the frictional parameters estimated by the PINN framework using a traditional numerical solver based on the Runge-Kutta method (dop853, Hairer et al. 1993), and comparing the results with the PINN output (Figure S1).

## Results and Discussion

To evaluate the short-term forecasting capability of the PINN-based DA framework, we compare the calculated surface displacements with GNSS observations under different DA periods. Figure 1b shows the displacement time series calculated using PINNs together with the observed GNSS time series for each DA window. In each panel, the black horizontal dotted line indicates the assimilation period; model results after this period shown by the red horizontal dotted line therefore correspond to forecasts based on the PINN estimation results. Regardless of the length of the assimilation data, all PINN-based models quantitatively reproduce the observed displacement time series during the assimilation period and achieve a significantly better fit than the results obtained by K24. This improvement is expected because the present PINN model incorporates spatially heterogeneous frictional properties, whereas K24 assumed spatially uniform friction (Fukushima et al. 2026). In contrast, the forecasted fault slip behavior differs substantially among the data periods. As demonstrated by K24, when DA is performed only up to the slip acceleration phase in a uniform friction model (e.g., data prior to 2010.5, Figure 1b(iv)-(vi)), the forecast tends to exhibit unstable fast slip unless prior constraints restricting the slip behavior to SSEs are imposed on the frictional parameters. In other words, it is essential to impose prior constraints to prohibit fast slip. In contrast, the PINN-based model forecasts slow transient slip even when observations corresponding only to slip acceleration are assimilated. This stabilization of fault slip behavior arises from the introduction of frictional heterogeneity, the physical implications of which are discussed below in terms of the estimated frictional parameters. In addition, the successful optimization of this high-dimensional inverse problem highlights the flexibility of NN representations, demonstrating that PINNs provide a powerful DA framework for fault slip monitoring and forecasting.

Figures 2−4 and S2 summarize the estimated frictional parameters, the calculated surface displacement, spatio-temporal evolution of fault slip velocity, and slip, respectively. Figure 2 shows that the frictional parameters are estimated on the order of $10^{-3}$ for *a*-*b* and *a*, and on the order of centimeters for *L*. The inferred spatial patterns of these parameters vary depending on the length of the assimilation period. When only data prior to the initiation of the SSE (i.e., before 2010.0, Figure 2(v)(vi) in each subpanel) are assimilated, PINN provides little information on frictional parameters. This outcome is expected because the GNSS observations during this period do not contain clear transient deformations (Figure 3(v)(vi)).

When data up to 2010.5 are assimilated, frictional heterogeneity with a strong velocity-weakening zone, i.e., large negative *a*-*b*, is clearly inferred beneath southwest Shikoku (Figure 3(iv)). In the model, SSE nucleation initiates in this region in early 2010 (Figure 4(iv) and S1(iv)) to explain the observed large displacement in Shikoku (Figure 3(iv)), consistent with previous studies (Yoshioka et al. 2015; Fukushima et al. 2026). This nucleation behavior can be interpreted as a consequence of velocity-weakening frictional properties and small *L* values in this region (Figure 2(iv) in each subpanel), which result in a small critical nucleation size $R_c$ (Equation 4), and thus promote frictional instability (Fukushima et al. 2026). After its initiation, slip propagates to the surrounding regions (Figure 4(iv)). In K24, forecasts based on assimilation up to 2010.5 resulted in unstable fast slip (Figure 1b(iv)-(vi)). Fault slip stability is controlled by the ratio between the radius of the SSE patch *R* and critical nucleation size $R_c$ (Chen and Lapusta 2009). In K24, only $R_c$ could vary because the SSE patch geometry was fixed, whereas in the present study both *R* and $R_c$ are allowed to vary through the introduction of frictional heterogeneity. As a result, the PINN-based forecasts exhibit stable slow slip behavior without imposing any prior constraints on the frictional parameters.

However, it should be noted that the improved performance is primarily attributable to the introduction of frictional heterogeneity rather than the intrinsic superiority of PINN framework itself. Similar results may also be obtained using other DA approaches that allow for spatially heterogeneous frictional parameters such as adjoint-based or ensemble methods. Nevertheless, estimating the spatial distribution of frictional parameters requires optimization in a high-dimensional parameter space. In this study, we adopted the PINN framework because it provides a flexible and efficient approach for handling such high-dimensional optimization problems by leveraging the strong representation ability of NNs.

For longer assimilation periods (up to 2011.0, 2011.5, and 2012.5), the estimated frictional parameters and slip velocity distributions become largely consistent across models (Figures 2(i)-(iii) in each subpanel and 4(i)-(iii)). In all cases, SSE nucleation occurs within the velocity-weakening patch beneath southwest Shikoku (Figure 3(i)-(iii)), followed by slip propagation into the surrounding regions. A notable difference among these models lies in the direction of slip propagation, which becomes increasingly restricted westward as the assimilation period becomes longer. This behavior reflects the emergence of velocity-strengthening regions that progressively surround the velocity-weakening zone (Figures 2(i)-(iii) in each subpanel and 4(i)-(iii)), thereby inhibiting slip propagation across these areas. To further examine this effect, we focus on the data misfit term $J_{\mathrm{data}}$ evaluated only over the period 2008.5–2010.5, during which the SSE initiates beneath southwest Shikoku and propagates westward. $J_{\mathrm{data}}$ takes almost the same values of ~1.53 for both cases with assimilation periods of up to 2010.5 and 2011.5, respectively. This indicates that the ability to explain the observed GNSS time series is comparable between these two models and that we cannot distinguish whether the slip propagates in a circular way or only westwards from the observations alone during 2008.5–2010.5. By assimilating observations extending after 2010.5, we can infer the subsequent slip west of the Bungo Channel to explain large displacement in Kyushu (Figures 3(i)-(iii) and 4). As a result, the direction of slip propagation is restricted to keep the mechanical consistency between the initial and the subsequent slips, leading to an updated shape of the velocity-weakening region as well as the appearance of the surrounding velocity-strengthening regions. These results suggest that PINN-based inversion mitigates the non-uniqueness of slip distributions by imposing fault mechanics as physical regularization. It should be noted that there are currently no independent geophysical or geological observations that directly confirm the existence of these velocity-strengthening regions. Therefore, their physical interpretation remains uncertain, and the inferred velocity-strengthening zones should be regarded as structures derived within the framework of the adopted crustal deformation data and physics-based governing equations.

More generally, the inferred frictional parameter distributions in Figure 2 should be interpreted as posterior estimates constrained by the observation data within the assimilation time window. Although the frictional parameters are modeled as intrinsic and time-independent properties, their inferred spatial distributions strongly depend on the sensitivity of the observation data to the model parameters. When the assimilated observations contain only stable steady slip behavior, the data provide weak constraints on the parameters (Figure 2(v)(vi) in each subpanel). Therefore, variations in the inferred parameter distributions with assimilation length should be interpreted not as temporal changes in the intrinsic frictional properties themselves, but rather as updates of their posterior estimates as additional observation constraints become available. In addition, the PINN framework introduces implicit regularization, which preferentially represents smooth slip distributions (Fukushima et al. 2025). As a result, short-wavelength slip variations are suppressed, reducing the tendency to overfit small temporal fluctuations in the observation data and mitigating possible aliasing associated with short-wavelength frictional heterogeneity. A more quantitative assessment of parameter uncertainty remains an important direction for future work, which will be discussed in the next section.

## Future Perspectives of fault slip monitoring and forecasting

We have demonstrated that the PINN-based DA approach substantially improves both the explanation of geodetic observations and the short-term forecasting of fault slip evolution compared to traditional DA methods such as K24. This improvement is primarily achieved by solving a high-dimensional optimization problem — namely, the estimation of spatially heterogeneous frictional parameters — by exploiting the flexible representation ability of NNs. The general framework of PINN-based fault slip monitoring and forecasting has now been largely established, and it can therefore be applied to other SSEs observed worldwide. More directly, frictional parameter distributions could be estimated by assimilating other SSEs that have occurred in the Bungo Channel, enabling systematic comparisons of similarities and differences among SSEs within the same framework. According to the kinematic inversion results (Seshimo and Yoshioka, 2022), the three SSEs that occurred in 2003, 2010, and 2019 exhibit similar total slip distributions. However, the 2003 and 2019 SSEs consist of two subevents, whereas the 2010 SSE does not. In particular, the second subevent of the 2019 SSE appears to resemble the 2010 SSE. Therefore, we expect similar frictional distributions in the region of the second subevent, while the frictional distributions associated with the first subevent may differ from those of the 2010 event. Assimilation of a series of multiple SSEs would be an important step toward examining whether a single model can consistently explain multiple SSE cycles. Such applications may provide a more realistic description of fault friction, including possible temporal variations of frictional properties associated with fluid migration, which can be naturally implemented within the current PINN framework.

Further improvement of the PINN-based DA framework is essential for more practical fault slip monitoring and forecasting. In the present study, we focus on point estimates and do not quantify their uncertainties. However, uncertainty estimates of frictional parameters are crucial for assessing their reliability and the temporal variability. Incorporating uncertainty quantification would further enable probabilistic short-term fault slip forecasting. Bayesian approaches, such as Bayesian PINNs (Yang et al. 2021), can incorporate uncertainty quantification into the PINN framework by treating NN parameters as probabilistic variables, and have already been applied to seismological problems (e.g., Agata et al. 2023). Introducing Bayesian formulations therefore represents a promising direction for advancing both fault physics understanding and practical fault slip forecasting.

Another important perspective concerns how strongly fault mechanics should be enforced in DA. Traditional DA methods developed for fault slip problems, such as adjoint method (Kano et al. 2015; 2020; Ohtani et al. 2025) and MCMC method (Fukuda et al. 2009; Kano et al. 2024), impose fault mechanics as hard constraints, such that the inferred fault slip evolution strictly satisfies the governing equations, i.e., $J_{\mathrm{ode}} = 0$ (Equation (7)), throughout the entire time period, except for numerical errors. In contrast, in the PINN-based DA framework (Fukushima et al. 2025; 2026), the governing equations are incorporated as weak constraints through loss functions for ordinary differential equations $J_{\mathrm{ode}}$, and thus mechanical consistency is not always guaranteed over the full data period. This distinction may be reasonable given that fault mechanics are still not fully understood. Although the reliability of fault mechanics remains under debate, several future directions can be considered within the PINN-based DA framework. One approach is to incorporate currently unmodeled physics. For example, the friction law in Equations (2) and (3) is an empirically derived constitutive law and therefore contains uncertainty regarding the underlying physics of fault friction. Modified versions of friction laws, such as those incorporating cut-off velocity effects (Shibazaki and Shimamoto 2007), may provide a more realistic description of SSE behavior. In addition, the governing equations adopted in this study do not explicitly include potentially important processes such as porosity evolution (Segall et al. 2010) and permeability evolution (Ozawa et al. 2024). Another approach is to incorporate PINN-based estimation of frictional parameters (NN_B in this study) into traditional DA methods. In this hybrid approach, fault mechanics would remain hard constraints, and fault slip evolution would be obtained by directly integrating the governing equations, while high-dimensional frictional parameter estimation would be achieved using NN. A further possibility is to explore unknown or unmodeled physical processes. While the specific form of such physics is currently unclear, observation data that cannot be fully explained by existing governing equations may contain information about missing physical processes (Fukushima et al. 2026), potentially leading to new physical insights — a key potential of PINN-based methods (Raissi et al. 2019; Chen et al. 2021). A first step in this direction would be to investigate, through numerical experiments, how PINNs behave when different fault physics, such as different evolution laws, are assumed.

In conclusion, we proposed a PINN-based DA framework for fault slip forecasting and applied it to the 2010 Bungo Channel SSE. By introducing frictional heterogeneity and solving a high-dimensional optimization problem using PINNs, we found that short-term forecasts yield slow transient slip evolution without imposing any prior constraint on fault friction, whereas traditional DA methods with uniform friction predict unstable fast slip. Through this application, we demonstrated the capability of PINNs for fault slip forecasting. The PINN-based DA framework thus represents a promising step toward physically consistent and data-adaptive fault slip monitoring and forecasting.

# Declarations

### Ethics approval and consent to participate

Not applicable.

### Consent for publication

Not applicable.

### List of abbreviations

SSE: slow slip events; DA: data assimilation; MCMC: Markov chain Monte Carlo; PINN: physics-informed neural network; GNSS: Global Navigation Satellite System; NN: neural networks; GEONET: GNSS earth observation network system

### Availability of data and materials

The original GNSS data from GEONET used in this study can be downloaded from https://www.gsi.go.jp/ENGLISH/geonet_english.html after registration (GSI 2025). Prof. Takuya Nishimura and Dr. Yusuke Tanaka kindly provided the preprocessed data (Tanaka and Nishimura 2024). The original PINN code can be downloaded from https://zenodo.org/records/14977672 (Fukushima 2025)

### Competing interests

The authors declare that they have no competing interests.

### Funding

This study was supported by the MEXT Project for Seismology toward Research Innovation with Data of Earthquake (STAR-E) (Grant number JPJ010217), the MEXT

Project for Seismology toward Research Innovation for Earthquake Evaluation with Exact Data (STAR-E NEXT) (Grant number JPJ013735) and JSPS KAKENHI (Grant numbers JP24H01019, JP24K02951, JP25K01084, JP25K01104).

## Authors' contributions

MK designed this study and conducted PINN-based DA. RF developed the PINN code. MK conducted the overall analysis and prepared the manuscript. All authors discussed the results and approved the final manuscript.

## Acknowledgements

We appreciate the two reviewers for their valuable comments on improving the manuscript and the editor Yuta Mitsui for handling the manuscript. Figures were generated using Generic Mapping Tools (Wessel et al. 2013).

## Authors' information

[1]Disaster Prevention Research Institute, Kyoto University, Gokasho, Uji, Kyoto 611-0011, Japan. [2]Japan Agency for Marine-Earth Science and Technology, 3173-25, Showa-cho, Kanazawa-ku, Yokohama, 236-0001, Japan. [3]Department of Geophysics, Stanford University, 397 Panama Mall, Stanford, CA, 94305, U.S.

## Figure legends

**Figure 1**. (a) The left panel shows the study area in southwest Japan, where the Philippine Sea Plate (PH) is subducting beneath the Amurian Plate (AM) along the Nankai Trough. The right panel shows the Bungo Channel region indicated by the black rectangle on the left panel. The red rectangle indicates the SSE fault region originally introduced by Hirahara and Nishikiori (2019) and adopted in subsequent studies (Kano et al. 2024; Fukushima et al. 2026). Green triangles represent GEONET stations used in this study. (b) Surface displacement time series at station 1059 shown in (a). Green, blue, and red dots indicate the daily GNSS time series in the trench-parallel (X), trench-perpendicular (Y), and vertical (Z) directions, respectively, while the red line represents the corresponding PINN-based calculations. The orange line indicates the calculations based on the estimation of Kano et al. (2024). The black and red horizontal dotted lines in each panel indicate the assimilation and forecast periods, respectively.

**Figure 2**. Spatial distribution of the estimated frictional parameters (a) $a$-$b(x,y)$, (b) $a(x,y)$, (c) $L(x,y)$, and (d) critical nucleation length $R_c(x,y)$. $R_c(x,y)$ is calculated only in the velocity-weakening region. In each subpanel, the end time of the assimilation period is shown.

**Figure 3**. Observed (black) and calculated (orange) surface displacements during the period from 2008.5 to the end of each data assimilation period. The corresponding spatial distributions of the estimated frictional parameter *a*-*b*(*x*,*y*) for each period are also shown. Values of variance reduction defined by $100 \times \left(1 - \mathbf{r}^{\mathrm{T}}\mathbf{r} / \mathbf{d}^{\mathrm{T}}\mathbf{d}\right)$ where **d** and **r** represent the observed displacement and the misfit between the observed and calculated vectors, respectively, are shown for reference.

**Figure 4**. Spatio-temporal evolution of fault slip velocity at the times shown at the top of each subpanel, with time measured from 2006.5 as 0.0 yr. In each panel, the end time of the assimilation period is shown in white.

## Figures

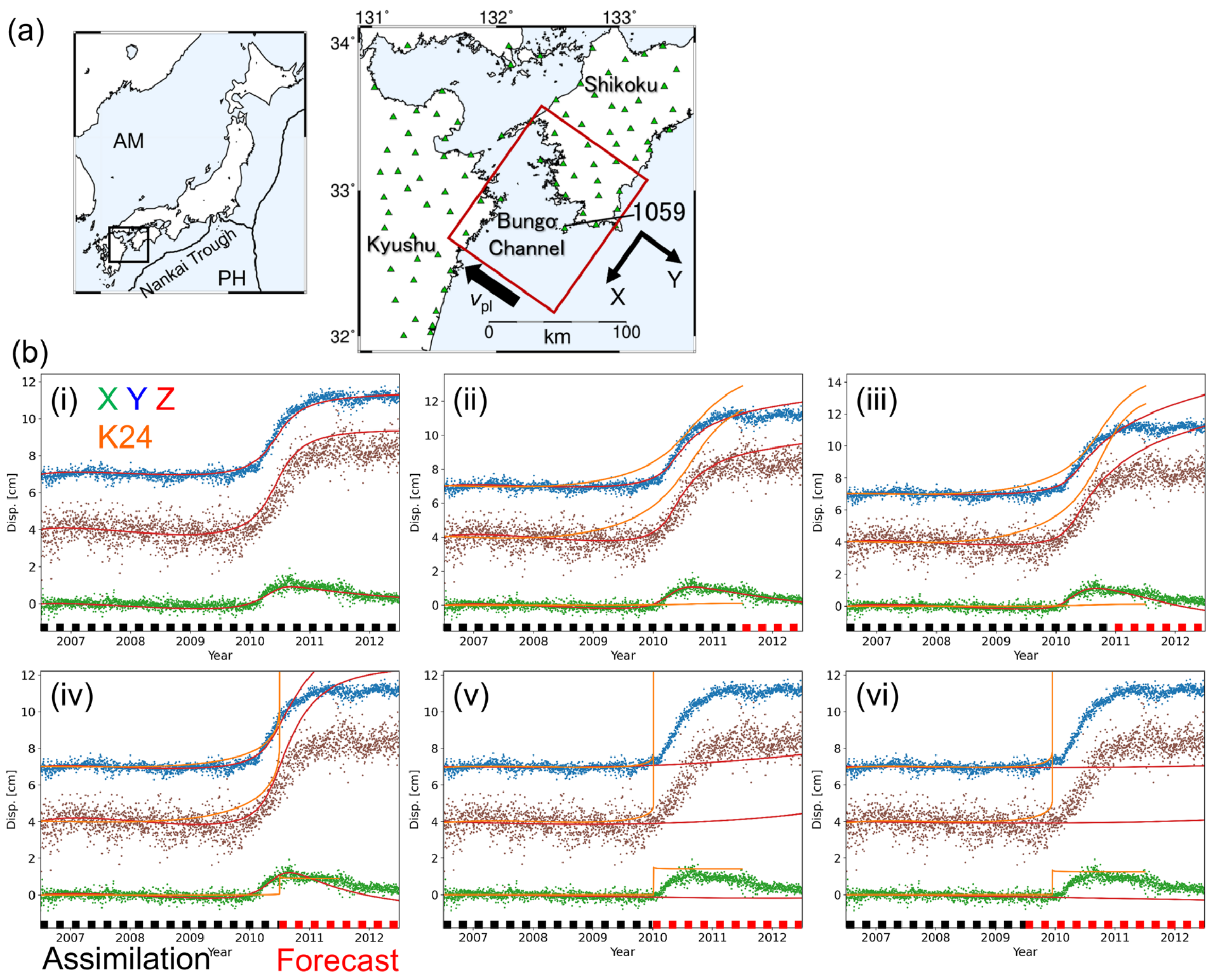

Figure 1

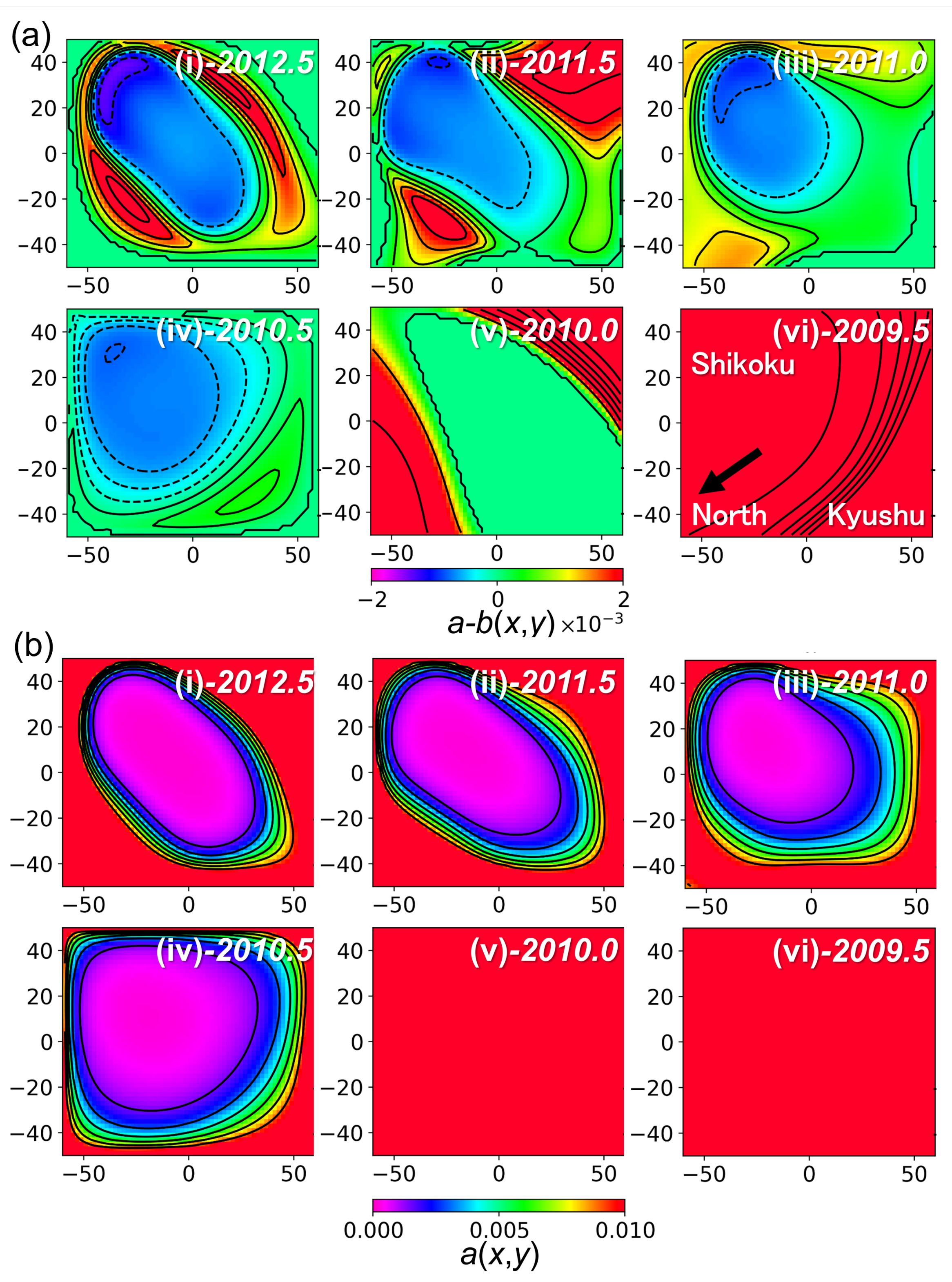

Figure 2

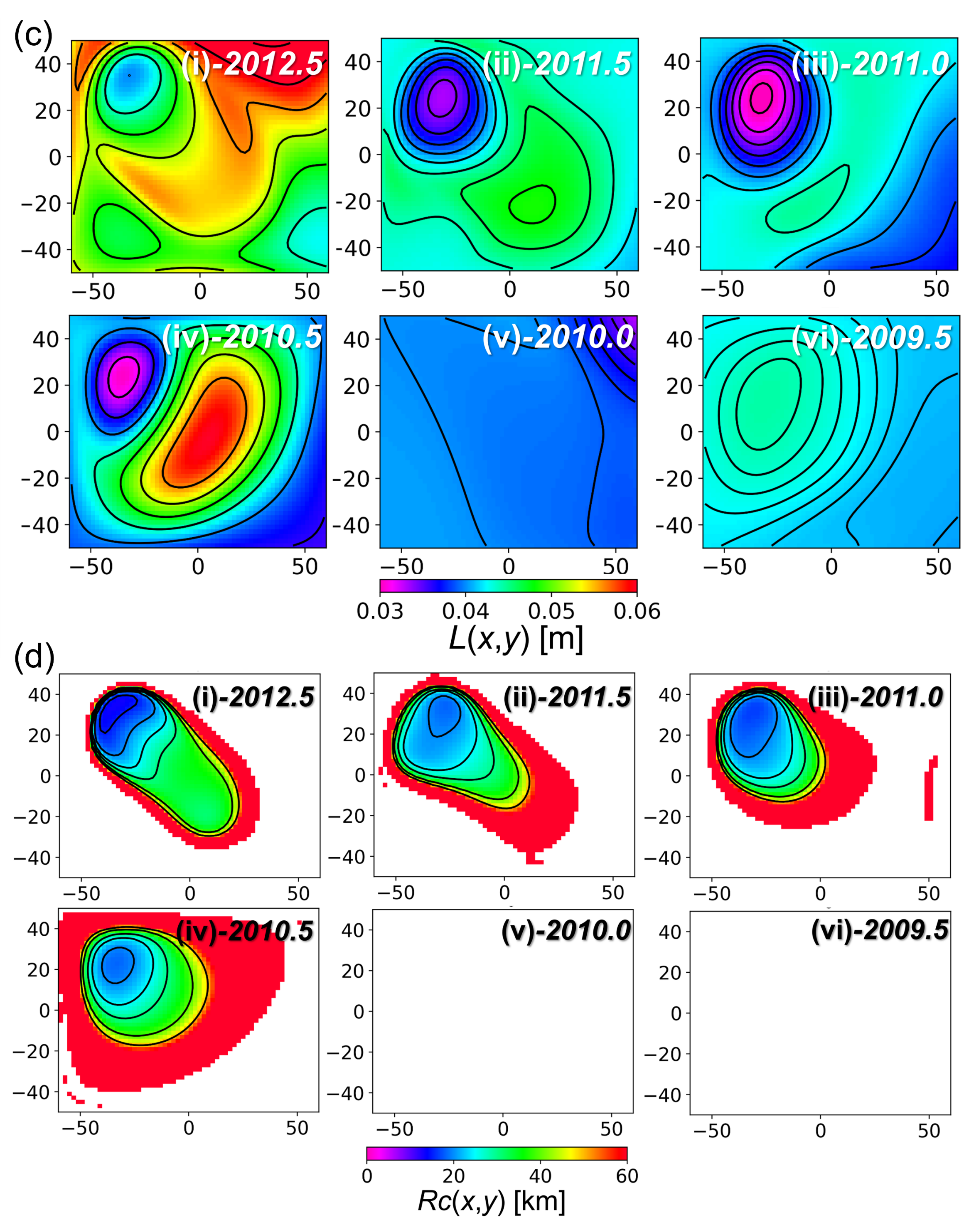

Figure 2 (continued)

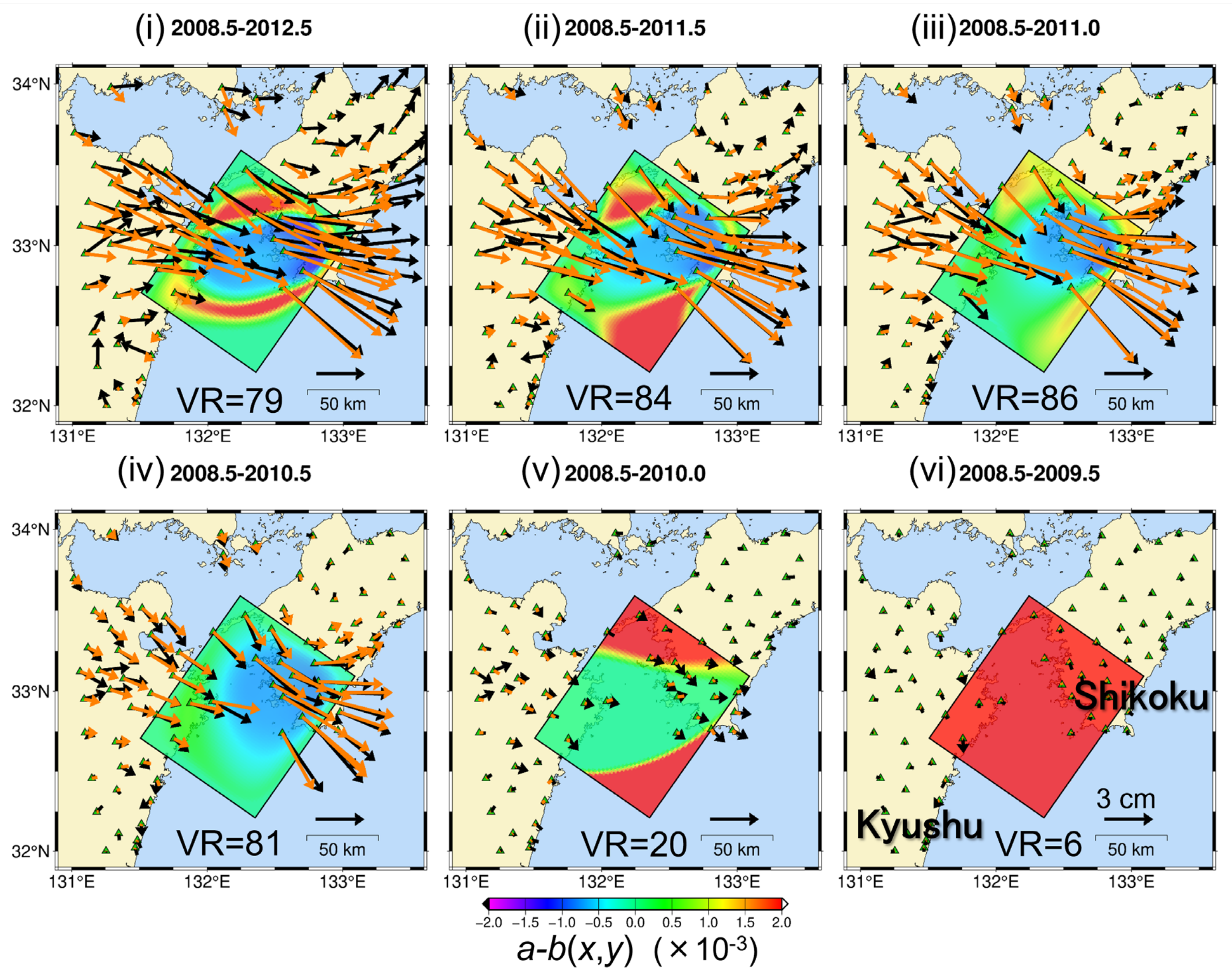

Figure 3

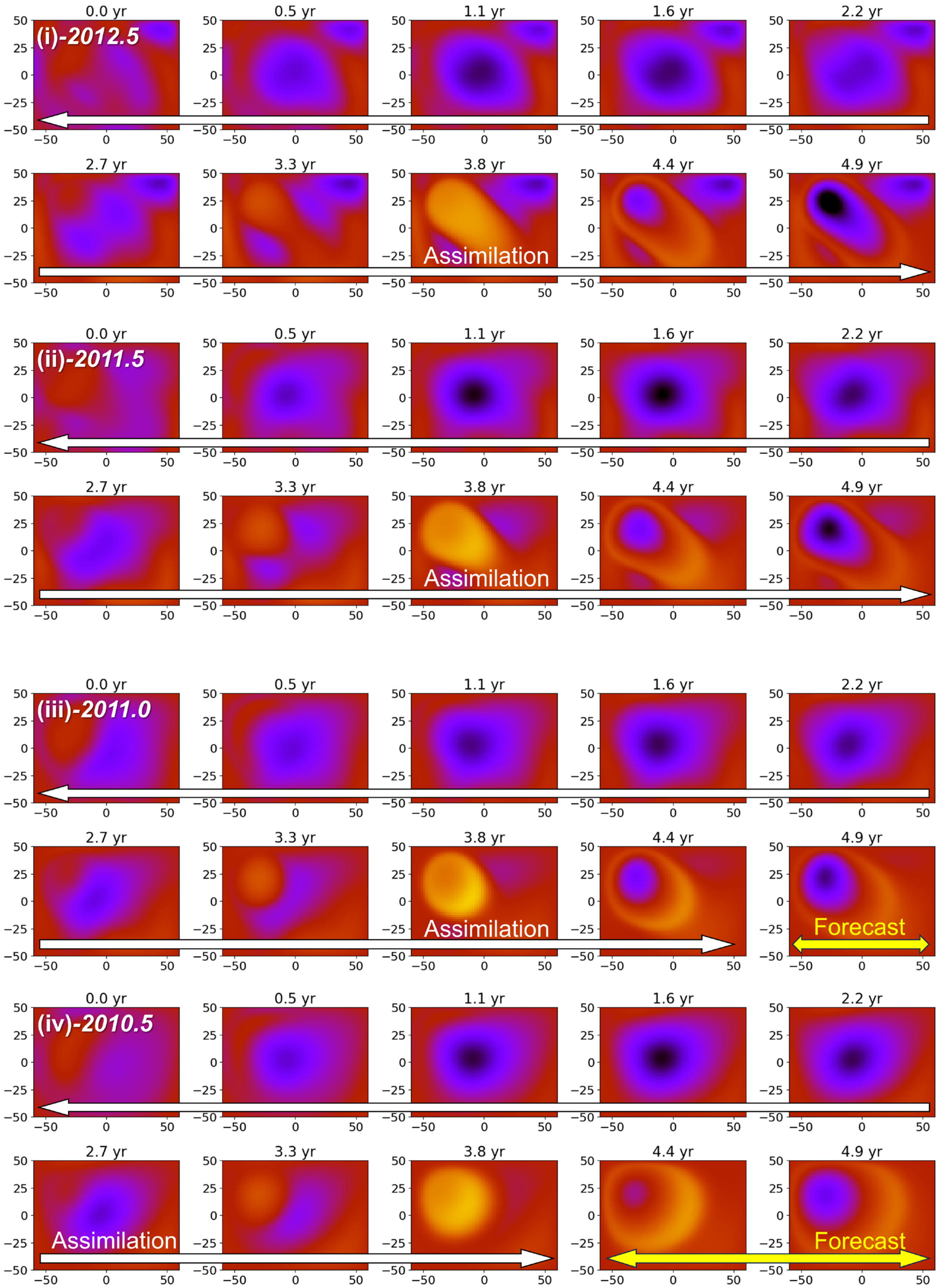

Figure 4

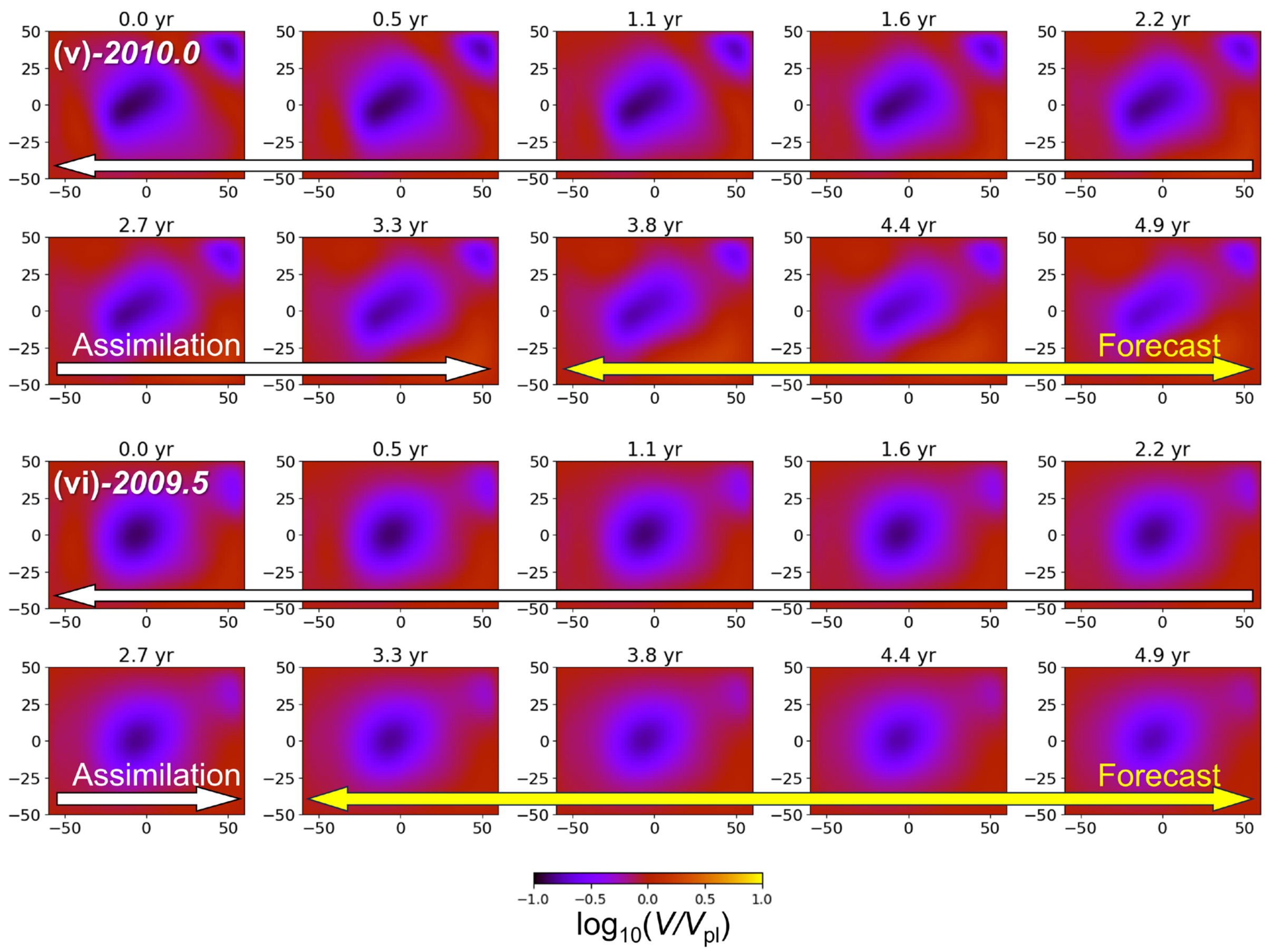

Figure 4 (continued)